\documentclass[aoas,preprint]{imsart}

\RequirePackage[OT1]{fontenc}
\RequirePackage{amsthm,amsmath}
\RequirePackage{natbib}
\RequirePackage[colorlinks,citecolor=blue,urlcolor=blue]{hyperref}

\pubyear{April 4, 2016}
\setattribute{journal}{name}{This is a preprint of an article published in Annals of Applied Statistics 10(2): 946--963 (2016)}

\startlocaldefs
\numberwithin{equation}{section}
\theoremstyle{plain}

\endlocaldefs

\RequirePackage{graphicx}

\begin{document}

\begin{frontmatter}

\title{Robust hyperparameter estimation protects against hypervariable genes and improves power to detect differential expression}
\runtitle{Robust empirical Bayes}

\author{\fnms{Belinda} \snm{Phipson}\thanksref{m1}\ead[label=e1]{belinda.phipson@mcri.edu.au}},
\author{\fnms{Stanley} \snm{Lee}\thanksref{m2,m3}},
\author{\fnms{Ian J.} \snm{Majewski}\thanksref{m2,m3}},
\author{\fnms{Warren S.} \snm{Alexander}\thanksref{m2,m3}},
\and
\author{\fnms{Gordon K.} \snm{Smyth}\corref{}\thanksref{m2,m3}\ead[label=e2]{smyth@wehi.edu.au}}

\affiliation{Murdoch Childrens Research Institute\thanksmark{m1}, The Walter and Eliza Hall Institute of Medical Research\thanksmark{m2} and The University of Melbourne\thanksmark{m3}}

\address{Murdoch Childrens Research Institute\\
50 Flemington Road\\
Parkville, 3052\\
Victoria, Australia\\
\printead{e1}}

\address{The Walter and Eliza Hall Institute of Medical Research\\
1G Royal Parade\\
Parkville, 3052\\
Victoria, Australia\\
\printead{e2}}

\runauthor{B. Phipson et al.}

\begin{abstract}
One of the most common analysis tasks in genomic research is to identify genes that are differentially expressed (DE) between experimental conditions.
Empirical Bayes (EB) statistical tests using moderated genewise variances have been very effective for this purpose, especially when the number of biological replicate samples is small.
The EB procedures can however be heavily influenced by a small number of genes with very large or very small variances.
This article improves the differential expression tests by robustifying the hyperparameter estimation procedure.
The robust procedure has the effect of decreasing the informativeness of the prior distribution for outlier genes while increasing its informativeness for other genes.
This effect has the double benefit of reducing the chance that hypervariable genes will be spuriously identified as DE while increasing statistical power for the main body of genes.
The robust EB algorithm is fast and numerically stable.
The procedure allows exact small-sample null distributions for the test statistics and reduces exactly to the original EB procedure when no outlier genes are present.
Simulations show that the robustified tests have similar performance to the original tests in the absence of outlier genes but have greater power and robustness when outliers are present.
The article includes case studies for which the robust method correctly identifies and downweights genes associated with hidden covariates and detects more genes likely to be scientifically relevant to the experimental conditions.
The new procedure is implemented in the limma software package freely available from the Bioconductor repository.
\end{abstract}


\end{frontmatter}

\section{Introduction}
\label{sec1}

Modern genomic technologies such as microarrays and RNA sequencing have made it routine for biological researchers to measure gene expression on a genome-wide scale.
Researchers are able to measure the expression level of every gene in the genome in any set of cells chosen for study under specified treatment conditions.
This article focuses on one of the most common analysis tasks, which is to identify genes that are differentially expressed (DE) between experimental conditions.

Gene expression experiments pose statistical challenges because the data are of extremely high dimension while the number of independent replicates of each treatment condition is often very small.
Simply applying univariate statistical methods to each gene in succession can produce imprecise results because of the small sample sizes.
Substantial gains in performance can be achieved by leveraging information from the entire dataset when making inference about each individual gene.

Empirical Bayes (EB) is a statistical technique that is able to borrow information in this way \citep{efron1973stein,morris1983parametric,casella1985empiricalbayes}.
EB has been applied very successfully in gene expression analyses to moderate the genewise variance estimators  \citep{baldi2001bayesian,wright2003random,smyth2004ebayes}.
These articles assume a conjugate gamma prior for the genewise variances and produce posterior variance estimates that are a compromise between a global variance estimate and individual genewise variance estimates.
The posterior variance estimators can be substituted in place of the classical estimators into linear model $t$-statistics and $F$-statistics.
\cite{wright2003random} and \cite{smyth2004ebayes} derived exact small sample distributions for the resulting moderated test statistics.
They showed that the EB statistics follow classical $t$ and $F$ distributions under the null hypothesis but with augmented degrees of freedom.
The additional degrees of freedom of the EB statistics relative to classical statistics represent the information that is indirectly borrowed from other genes when making inference about each individual gene.

EB assumes a Bayesian hierarchical model for the genewise variances but, instead of basing the prior distribution on prior knowledge as a Bayesian procedure would do, the prior distribution is estimated from the marginal distribution of the observed data.
\cite{smyth2004ebayes} developed closed-form estimators for the parameters of the prior distribution from the marginal distribution of the residual sample variances.
This procedure is implemented in the limma software package \citep{ritchie2015limma} and the resulting EB tests have been shown to offer improved statistical power and false discovery rate (FDR) control relative to the ordinary genewise $t$-tests, especially when the sample sizes are small \citep{kooperberg2005significance,murie2009comparison,ji2010analyzing,jeanmougin2010should}.
The limma software has been used successfully in thousands of published biological studies using data from a variety of genomic technologies, especially studies using expression microarrays and RNA-seq.

\begin{figure}
\begin{center}
\includegraphics[height=2.35in,width=4.9in]{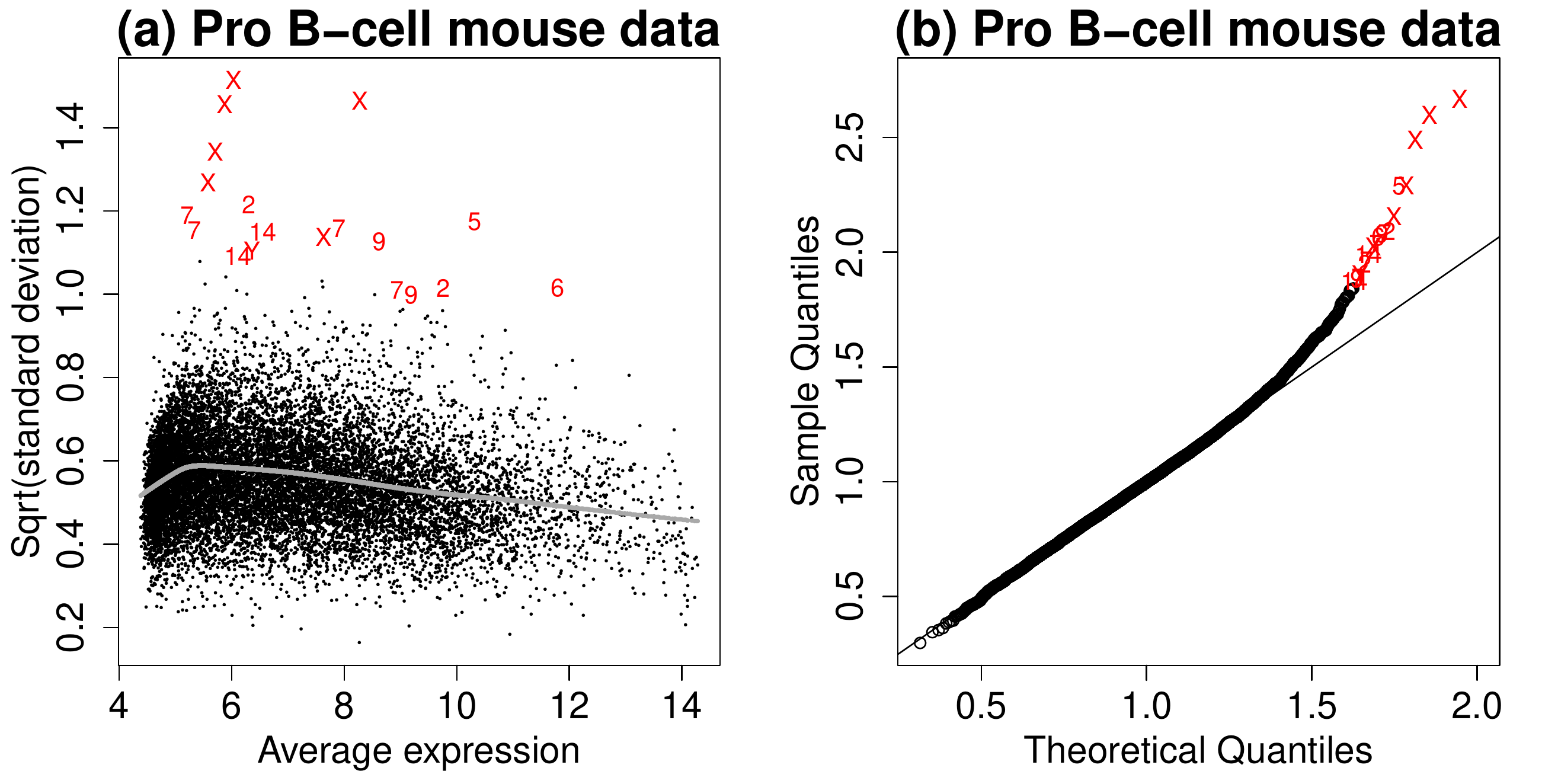}
\includegraphics[height=2.35in,width=4.9in]{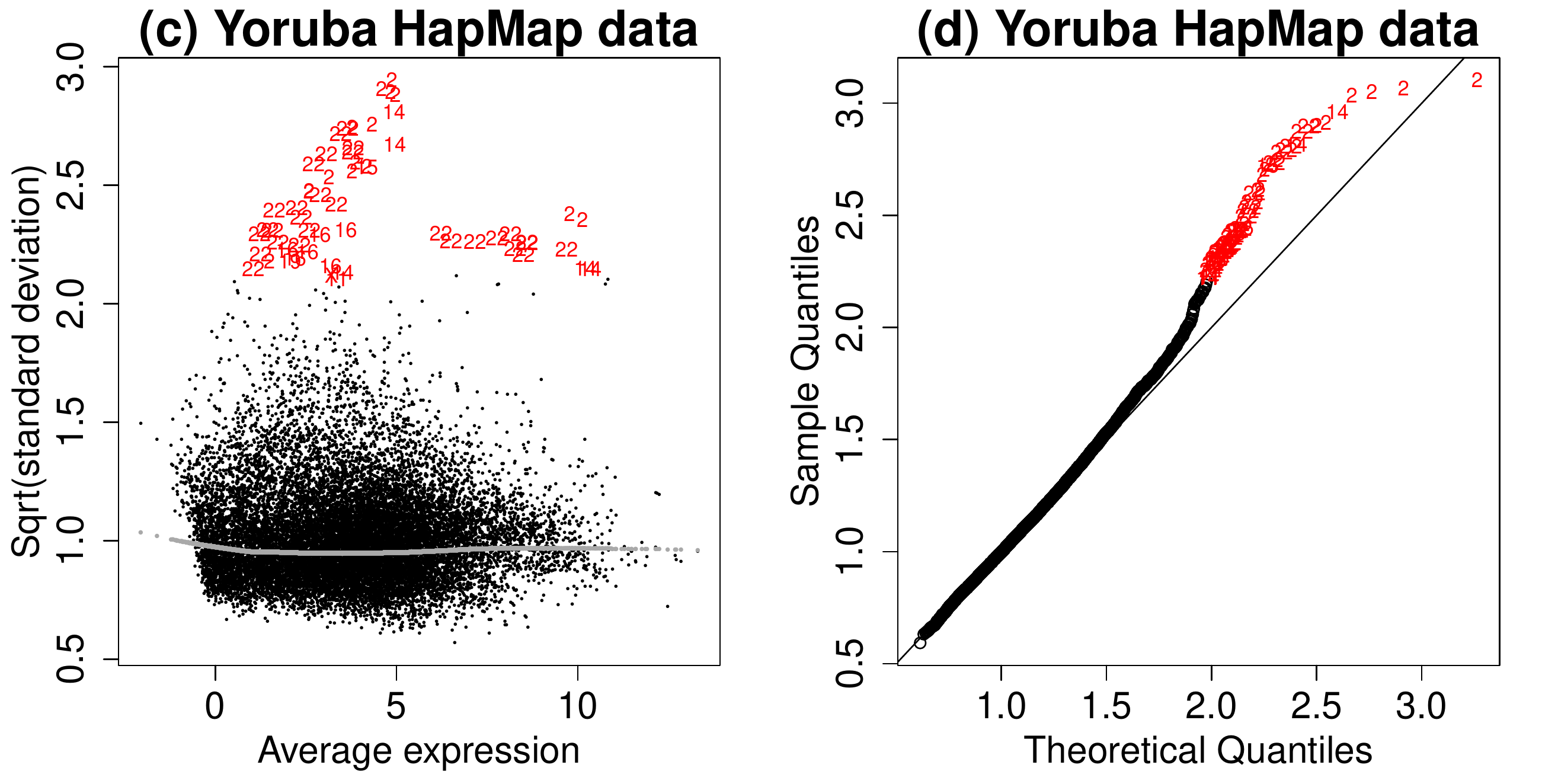}
\includegraphics[height=2.35in,width=4.9in]{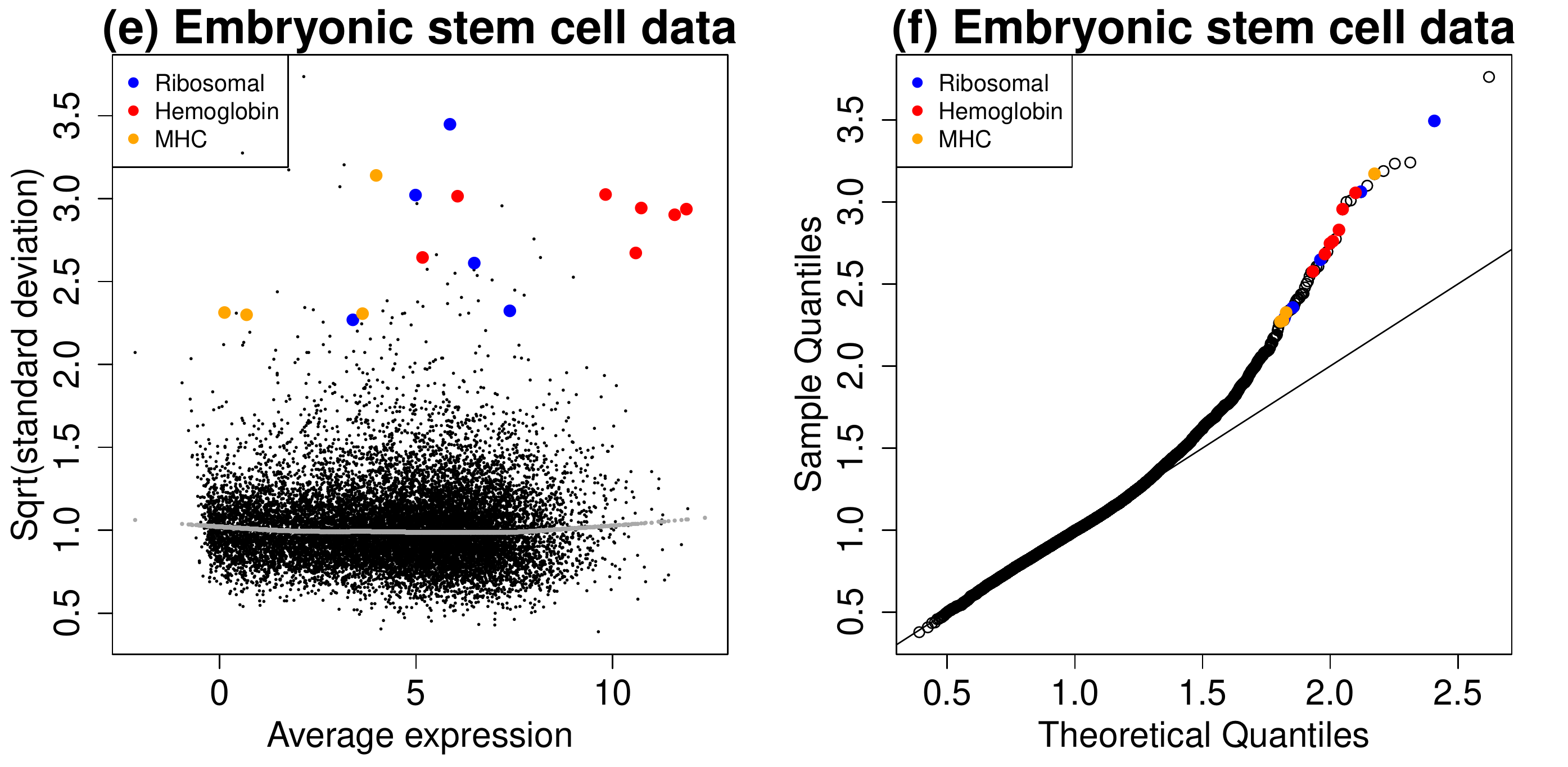}
\end{center}
\caption{
Hypervariable genes for the three case study datasets.
Left panels plot genewise residual standard deviations vs average log2-expression.
Right panels show probability plots of the standard deviations after transforming to equivalent normal deviates.
Hypervariable genes are marked by chromosome number in panels a--d and by gene function in panels e--f.
For each dataset the hypervariable genes have clear biological interpretations.
}
\label{PlotsShowingOutliers}
\end{figure}

This article improves the limma EB differential expression tests by robustifying the hyperparameter estimation procedure.
As in the original method, we fit genewise linear models to the log-expression values and extract residual variances, but now we give special attention to residual variances that are exceptionally large or exceptionally small.
Genes corresponding to extreme variances will be considered `outliers'.
Following terminology used in the genomics literature, we refer to outlier genes with large variances as `hypervariable genes'.
We show that, for certain genomic datasets, a small number of outlier genes can have an undesirable influence on the hyperparameter estimators, decreasing the effectiveness of the EB differential expression procedures.
(Figure~\ref{PlotsShowingOutliers} plots residual standard deviations for three datasets.
In each study there are special biological factors that cause a subset of the genes to have larger than expected residual variances.)
We show that in such cases the effectiveness of EB can be restored by a robust approach that isolates the outlier genes.
We develop a robust estimation scheme with a positive breakdown point for the hyperparameters and incorporate this into the differential expression procedure.
The robust EB procedure has the effect of decreasing the informativeness of the prior distribution for hypervariable genes while increasing its informativeness for other genes.
This effect has the double benefit of reducing the chance that hypervariable genes will be spuriously identified as DE while increasing statistical power for the main body of genes.

Our robust EB approach uses more diffuse prior distributions for the variances of hypervariable genes.
A conjugate prior is still used for each gene and this allows us to preserve a key feature of the original EB differential expression procedures, which is the ability to derive exact small-sample null distributions for the test statistics.
Our robust EB approach is fast and numerically stable without difficult convergence issues.
It reduces exactly to the original EB procedure when there are no hypervariable genes.
Simulation studies show that the robustified tests for differential expression have similar performance to the original tests in the absence of outlier genes but have greater power and robustness when they are present.

To the best of our knowledge there has been no previous work on robust EB for variances.
Most robust EB schemes in other contexts have been based on heavy-tailed prior distributions.
We have avoided such an approach because crucial advantages of the original DE procedures would thereby be lost, in particular the posterior mean variance estimators would no longer be available in closed form and the test statistics would no longer yield exact p-values.
\cite{efron1972limiting} proposed limited translation rules when estimating means of standard normal distributions.
This proposal originated the idea of limited learning for extreme cases, but with the aim of limiting the bias in extreme cases rather than improving estimation of the hyperparameters.
\cite{gaver1987robust} analyzed a Poisson process using a heavy-tailed (log-Student) prior distribution for the Poisson mean.
This approach achieves insensitivity to outliers but loses efficiency as well as being less mathematically tractable.
\cite{liao2014prior} estimated log-fold expression changes and achieved robustness with respect to misspecified working priors by conditioning on the rank of each estimated log-fold-change rather than on the actual observation.
Again this is less mathematically tractable than our approach.

Our robustified EB procedure has been implemented in the limma software package and can be invoked by the option \texttt{robust=TRUE} in calls to the \texttt{eBayes} or \texttt{treat} \citep{mccarthy2009treat} functions.
Invoking the option requires no other changes to the analysis pipelines from a user point of view.
All downstream functions recognize and work with robust EB results as appropriate.

The following sections review the EB approach to differential expression, then derive the robust estimators and modified differential expression scheme.
We evaluate the performance of the robustified procedure using simulations, then present three case studies in which the EB approach identifies genetic instabilities specific to each study.
We give a detailed analysis of a microarray study of PRC2 function in pro-B cells for which a gender effect produces hypervariable genes.
The EB procedure is effective at downweighting sex-linked genes associated with the unwanted covariate in favor of genes of more scientific interest.
Finally we discuss possible generalizations of our robust EB approach to other contexts.

\section{Linear models and moderated $t$-statistics}

Consider a genomic experiment in which the expression levels of $G$ genes are measured for $n$ RNA samples.
We follow the notation and linear model formulation introduced by \cite{smyth2004ebayes}.
Write $y_{gi}$ for the log-expression level of gene $g$ in sample $i$.
The log-expression values satisfy genewise linear models
\[
E(y_g)=X\beta_g
\]
where $y_g$ is the column vector $(y_{g1},\ldots,y_{gn})^T$, $X$ is an $n\times p$ design matrix of full column rank representing the experimental design and $\beta_g$ is an unknown coefficient vector that parametrizes the average expression levels in each experimental condition.
For each gene, the $y_{gi}$ are assumed independent with
\[
\textrm{var}(y_{gi})=\sigma_g^2/w_{gi},
\]
where $\sigma_g^2$ is an unknown variance and the $w_{gi}$ are known weights.
The least squares coefficient estimator is
\[
\hat{\beta}_g = (X^TW_gX)^{-1}X^TW_gy_g
\]
where $W_g$ is the diagonal matrix with elements $w_{g1},\ldots,w_{gn}$.
The residual sample variances are
\[
s^2_g=(y_g-\hat\mu_g)^T(y_g-\hat\mu_g)/d_g
\]
where $\hat\mu_g=X\hat\beta_g$ and $d_g$ is the residual degrees of freedom.
Usually $d_g=n-p$, but genes with missing $y$ values or zero weights may have smaller values for $d_g$.
Conditional on $\sigma^2_g$, $d_gs^2_g/\sigma^2_g$ is assumed to follow a chisquare distribution with $d_g$ degrees of freedom,
an assumption we write as
\[
s^2_g|\sigma^2_g\sim \sigma_g^2 \chi^2_{d_g}/d_g.
\]

We assume a conjugate prior distribution for $\sigma^2_g$ in order to stabilize the genewise estimators.
The $\sigma_g^2$ are assumed to be sampled from a scaled inverse chi-square prior distribution with degrees of freedom $d_0$ and location $s_0^2$,
\[
\sigma^2_g \sim s_0^2 d_0/\chi^2_{d_0}.
\]
It follows that the posterior distribution of $\sigma^2_g$ given $s^2_g$ is scaled inverse chi-square,
\[
\sigma_g^2|s_g^2 \sim \frac{d_0s_0^2 + d_gs_g^2}{\chi_{d_0+d_g}^2}.
\]
and the posterior expectation of $1/\sigma_g^2$ given $s_g^2$ is $1/\tilde s_g^2$ with
\[
\tilde s_g^2 = \frac{d_0s_0^2+d_gs_g^2}{d_0+d_g}
\]
The $\tilde s_g^2$ are the EB moderated variance estimators.
The moderated $t$-statistic for a given coefficient $\beta_{ig}$ is
\[
\tilde t_{gj} = \frac{\hat \beta_{gj}}{\tilde s_g \sqrt{v_i}}
\]
where $v_i$ is the $i^{th}$ diagonal element of $(X^TW_gX)^{-1}$.
If the null hypothesis $\beta_{gj}=0$ is true, then $\tilde t_{gj}$ follows a $t$-distribution on $d_g+d_0$ degrees of freedom \citep{smyth2004ebayes}.
In general, any ordinary genewise $t$ or $F$-statistic derived from the linear model can be converted into an EB moderated statistic by substituting $\tilde s^2_g$ for $s^2_g$, in which case the denominator degrees of freedom for the null distribution increase from $d_g$ to $d_0+d_g$.

\section{Robust hyperparameter estimation}
\label{section:robest}

Under the above hierarchical model, $s^2_g$ follows a scaled $F$-distribution on $d_g$ and $d_0$ degrees of freedom,
\[
s^2_g \sim s_0^2 F_{d_g,d_0}.
\]
The log-variances $\log s^2_g$ follow Fisher's z-distribution, which is roughly symmetric and has finite moments of all orders.
The limma package estimates the hyperparameters $s^2_0$ and $d_0$ by matching the theoretical mean and variance of the z-distribution to the observed sample mean and variance of the $z_g$.
The empirical estimates $s^2_0$ and $d_0$ are then substituted into the above formulas to obtain $\tilde t_{gj}$ and to conduct genewise statistical tests for differential expression.

As the observed variance of the $\log s^2_g$ increases, the estimated value of $d_0$ decreases, meaning that less information is borrowed from the prior to form the moderated $t$-statistics.
In the examples shown in Figure~\ref{PlotsShowingOutliers}, the variance of $\log s^2_g$ would be much reduced if a small number of the most variable genes were excluded.
We therefore seek to replace limma's moment estimation scheme for the hyperparameters with a robust version.

Our approach is to apply moment estimation to the Winsorized sample variances.
The idea of Winsorizing is to reset a specified proportion of the most extreme sample variances to less extreme values \citep{tukey1962future}.
Let $p_u$ and $p_l$ be the maximum proportion of outliers allowed in the upper and lower tails of the $s^2_g$ respectively.
Typical values are $p_l=0.05$ and $p_u=0.1$, although any values strictly between 0 and 0.5 are permissable.
Let $q_l$ and $q_u$ be the corresponding quantiles of the empirical distribution of $s_g^2$, so that $p_l$ of the variances are less than or equal to $q_l$ and $p_u$ are greater than or equal to $q_u$.
The empirical Winsorizing transformation is defined by
$$
{\rm win}(s_g^2) = \left\{ \begin{array}{ll}
q_l & \textrm{if } s_g^2 \leq q_l\\
s_g^2 & \textrm{otherwise}\\
q_u & \textrm{if }  s_g^2 \geq q_u.
\end{array} \right.
$$
Write $z_g=\log {\rm win}(s_g^2)$ for log-transformed Winsorized variances, and
let $\bar z$ and $s_z^2$ be the mean and variance of the observed values of $z_g$.

Define the Winsorized $F$-distribution as follows.
If $f \sim F_{d_g,d_0}$ then the Winsorized random variable is
$$
{\rm win}(f) = \left\{ \begin{array}{ll}
q_l & \textrm{if } f \leq q_l\\
f & \textrm{otherwise}\\
q_u & \textrm{if }  f \geq q_u.
\end{array} \right.
$$
where now $q_l$ and $q_u$ are the lower tail $p_l$ and upper tail $p_u$ quantiles of the $F_{d_g,d_0}$ distribution.

Write $\nu(d_g,d_0)$ and $\phi(d_g,d_0)$ for the expected value and variance of $\log {\rm win}(f)$.
An efficient and accurate algorithm for computing $\nu$ and $\phi$ using Gaussian quadrature is described in Section~\ref{supp}.

Assuming that the $d_g$ are all equal, the hyperparameter $d_0$ is estimated by equating $s^2_z=\phi(d_g,d_0)$ and solving for $d_0$ using a modified Newton algorithm \citep{brent1973algorithms}.
Having estimated $d_0$, the logarithm of the parameter $s^2_0$ is estimated by $\bar z - \nu(d_g,\hat d_0)$.

If the $d_g$ are not all equal, then the $s^2_g$ are transformed to equivalent random variables with equal $d_g$ before applying the above algorithm.
Details of this transformation are given in Section~\ref{supp}.

\section{Gene-specific prior degrees of freedom}

Having estimated $d_0$ and $s_0$ robustly, we can identify genes that are outliers in that their variances are too large to have reasonably arisen from the estimated prior.
The question naturally arises as to how to handle such outliers.
One reasonable approach would be to ignore the prior information for such genes, on the basis that the prior appears to be inappropriate.
This approach would assign $d_0=0$ for such genes, meaning that ordinary $t$-tests would be used for these genes instead of EB moderated statistics.
On the other hand, the prior should in principle still have some limited relevance even for the outliers.
Our approach is to assign gene-specific prior degrees of freedom, $d_{0g}$, whereby $d_{0g}=d_0$ for non-outlier genes but outlier genes are assigned smaller values depending on how extreme the outlier is.
In effect we assume that each hypervariable gene $g$ has a true variance $\sigma^2_g$ sampled from $s_0^2 d_{0g}/\chi^2_{d_{0g}}$ with $0<d_{0g}<d_0$.

We start by identifying a lower bound for the $d_{0g}$ from the largest observed $s^2_g$ value.
Specifically we find $d_{\rm outlier}$ such that the maximum $s^2_g$ is equal to the median of the $s_0^2 F_{d_g,d_{\rm outlier}}$ distribution.
A fast stable numerical algorithm for finding $d_{\rm outlier}$ is given in Section~\ref{supp}.

Next we evaluate the posterior probability that each gene is a hypervariable outlier.
Let $p_g$ be the $p$-value for testing whether gene $g$ is a outlier, defined by $p_g=P(f>s^2_g/s^2_0)$ where $f\sim F_{d_g,d_0}$.
Let $\pi_0$ be the prior probability that gene $g$ is not an outlier and let $r_g$ be the marginal probability of observing a residual variance more extreme than $s^2_g$.
The posterior probability given $s^2_g$ that case $g$ is not an outlier is
$\pi_g=p_g \pi_0 /r_g$.
Assuming that most genes are not outliers, we conservatively set $\pi_0=1$.
The marginal probability $r_q$ can be estimated empirically from the rank of $s^2_g$ amongst all the observed values of $s^2_g$, 
i.e., by $r_g=(r-0.5)/G$ where $r$ is the rank of $s^2_g$.
Substituting these values in the above formula yields a conservative estimate $\pi_g=p_g/r_g$.

The initial estimate of $\pi_g$ is not necessarily monotonic in $s^2_g$ or $p_g$.
We ensure that $\pi_g$ is a non-decreasing function of $p_g$ in the following manner.
First the cases are ordered in increasing order of $p_g$.
Then the cumulative mean $\bar \pi_g=(1/g)\sum_{i=1}^g\pi_i$ is computed for each $g$.
Let $g_m$ be the first value of $g$ for which $\bar \pi_g$ achieves its minimum.
All $\pi_g$ for $g=1,\ldots,g_m$ are set to the minimum value of $\bar\pi_g$,
to allow for the possibility that $\pi_g$ might be small for a group of cases but not for the most extreme case.
Finally, a cumulative maximum filter is applied to the $\pi_g$, after which the $\pi_g$ are non-decreasing.
In practice, this process is nearly identical to using isotonic regression \citep{barlow1972isotonic} to enforce monotonicity.

Finally the genewise prior degrees of freedom are defined by
$$d_{0g}=\pi_g d_0+(1-\pi_g) d_{\rm outlier}.$$
This process ensures that $d_{0g}=d_0$ for most genes, but any gene that is a clear outlier with a very small $p_g$ value will be assigned a much lower value.

\section{Covariate dependent priors}
\label{section:covdep}

In gene expression experiments, the variance of the log-expression values often depends partly on the magnitude of the expression level \citep{sartor2006intensity,law2014voom}.
It is therefore helpful to extend the EB principle to permit the prior variance $s^2_0$ to depend on the average log-expression $A_g$ of each gene.
This extension generalizes the prior distribution for $\sigma^2_g$ to be gene-specific:
$$\sigma^2_g \sim s^2_{0g} \chi^2_{d_0}/d_0$$
where $s^2_{0g}$ varies smoothly with $A_g$.
In other words, the prior distribution depends on the covariate $A_g$.
Such a strategy is implemented in the limma package \citep{ritchie2015limma}.

Our strategy for robust EB with a variance trend is as follows.
First we fit a robust lowess trend \citep{cleveland1979lowess} to $\log s^2_g$ as a function of $A_g$.
We detrend the $\log s^2_g$ by subtracting the fitted lowess curve, then apply the robust EB algorithm described above to the detrended variances.
The final genewise prior values $s^2_{0g}$ are the product of the unlogged lowess trend and the $s^2_0$ estimated from the detrended variances.

\section{Software implementation}
\label{section:implementation}

The robust hyperparameter estimation strategy described above is implemented in the limma function \texttt{fitFDist\-Robustly}.
The tail proportions for Winsorizing are user settable with defaults $p_l=0.05$ and $p_u=0.1$.
This function is called by \texttt{squeezeVar}, which computes EB moderated variances.
\texttt{squeezeVar} in turn is called by user-level functions including \texttt{eBayes} and \texttt{treat} in the limma package and \texttt{estimateDisp} and \texttt{glmQLFit} in the edgeR package \citep{robinson2010edger}.
These functions integrate the robust EB strategy into analysis pipelines for gene expression microarrays and RNA-seq.
\texttt{glmQLFit} is also called in analysis pipelines of the csaw and diffHic packages for the analysis of ChIP-seq and Hi-C sequencing data \citep{lun2015csaw,lun2015diffhic}.

\begin{figure}
\begin{center}
\includegraphics[height=2.35in,width=2.3in]{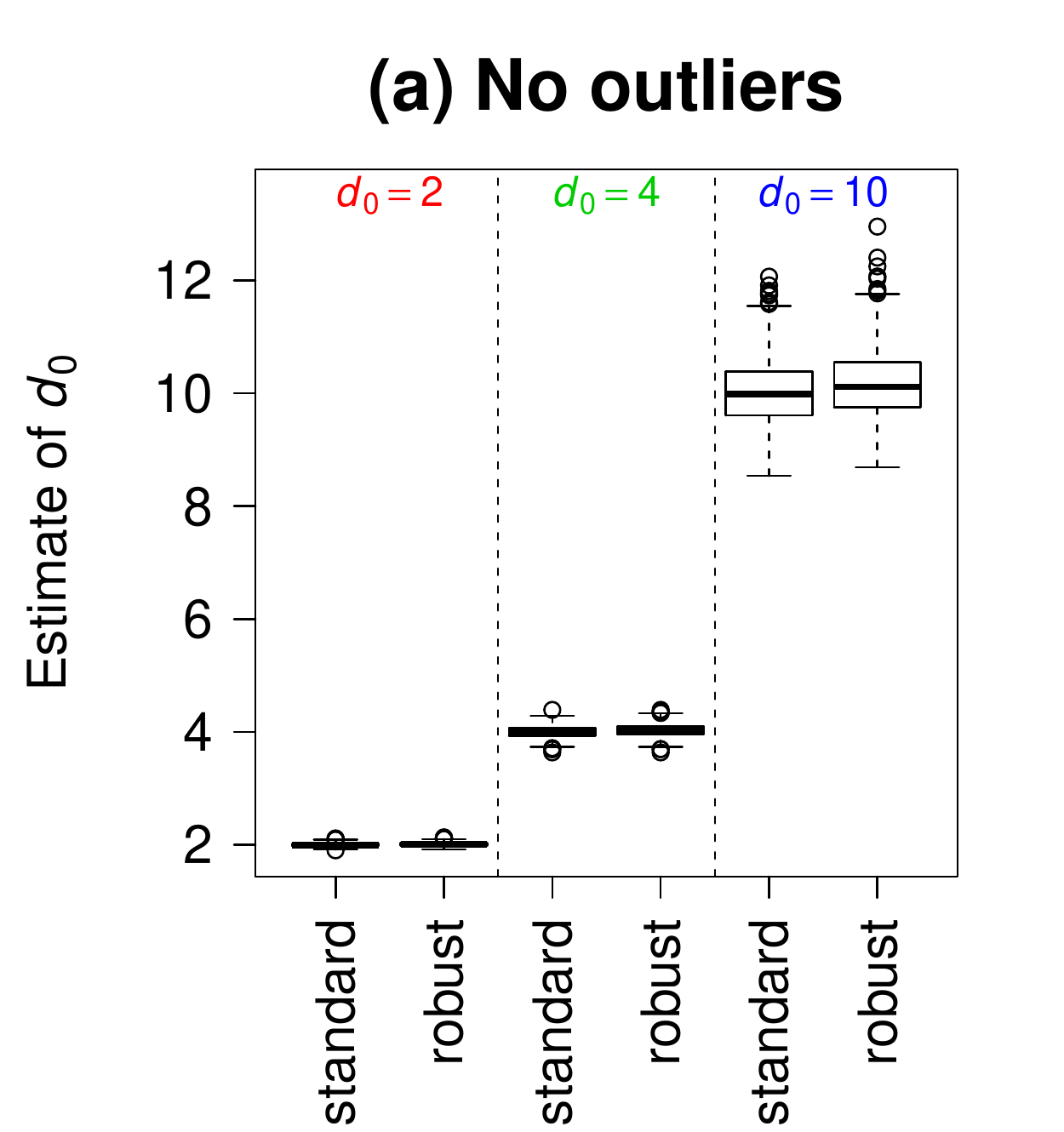}
\includegraphics[height=2.35in,width=2.3in]{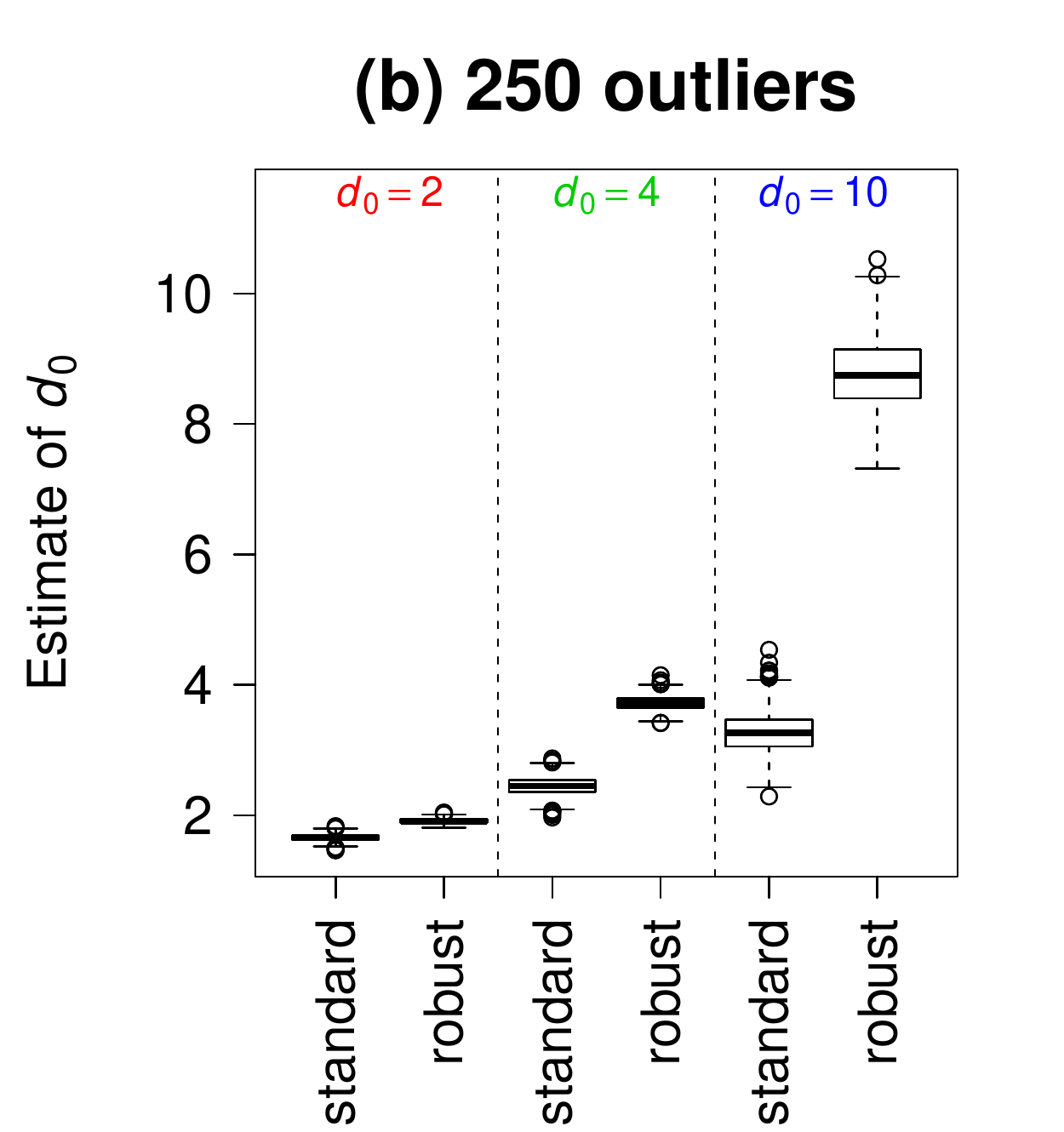}
\includegraphics[height=2.35in,width=2.3in]{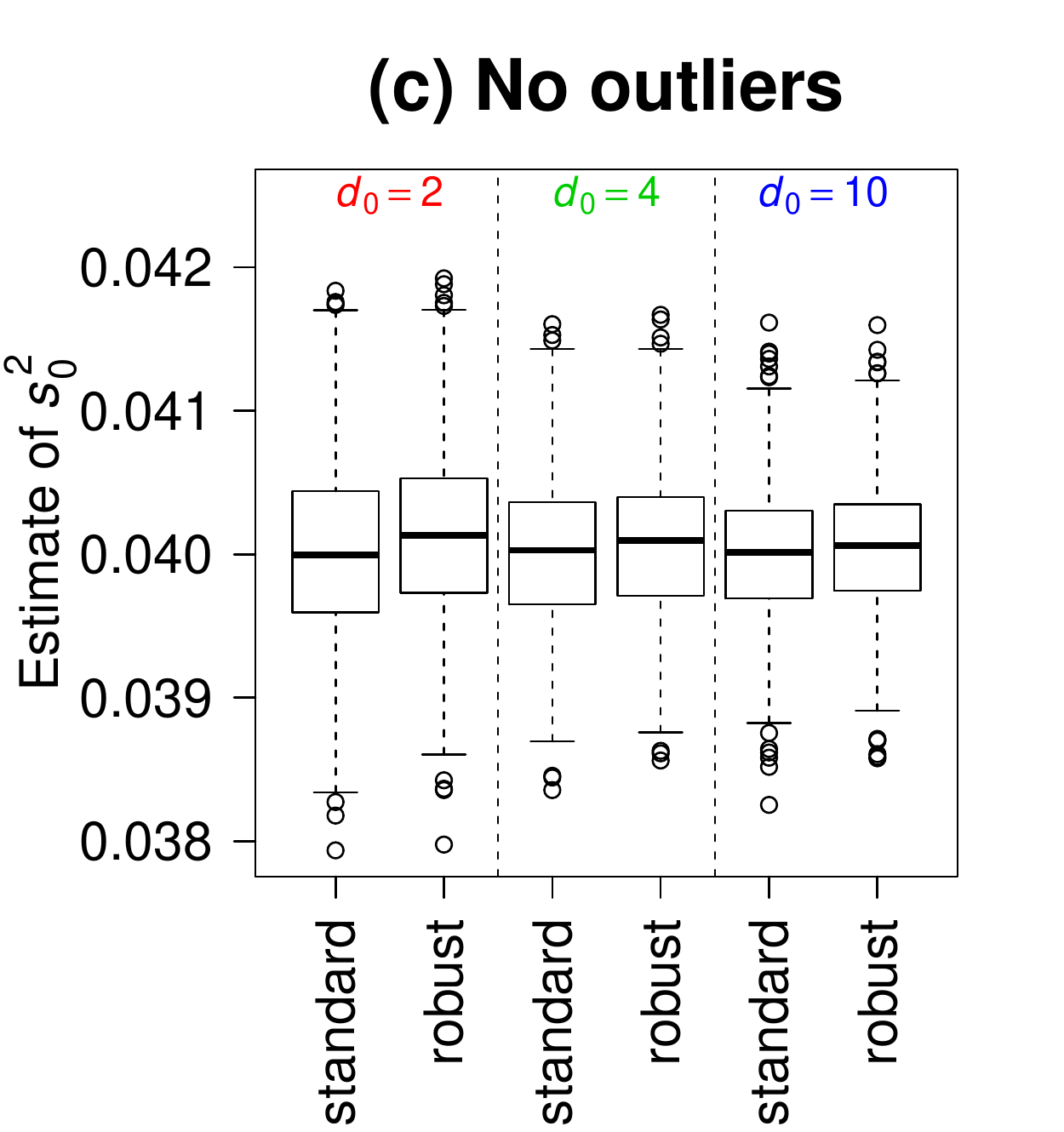}
\includegraphics[height=2.35in,width=2.3in]{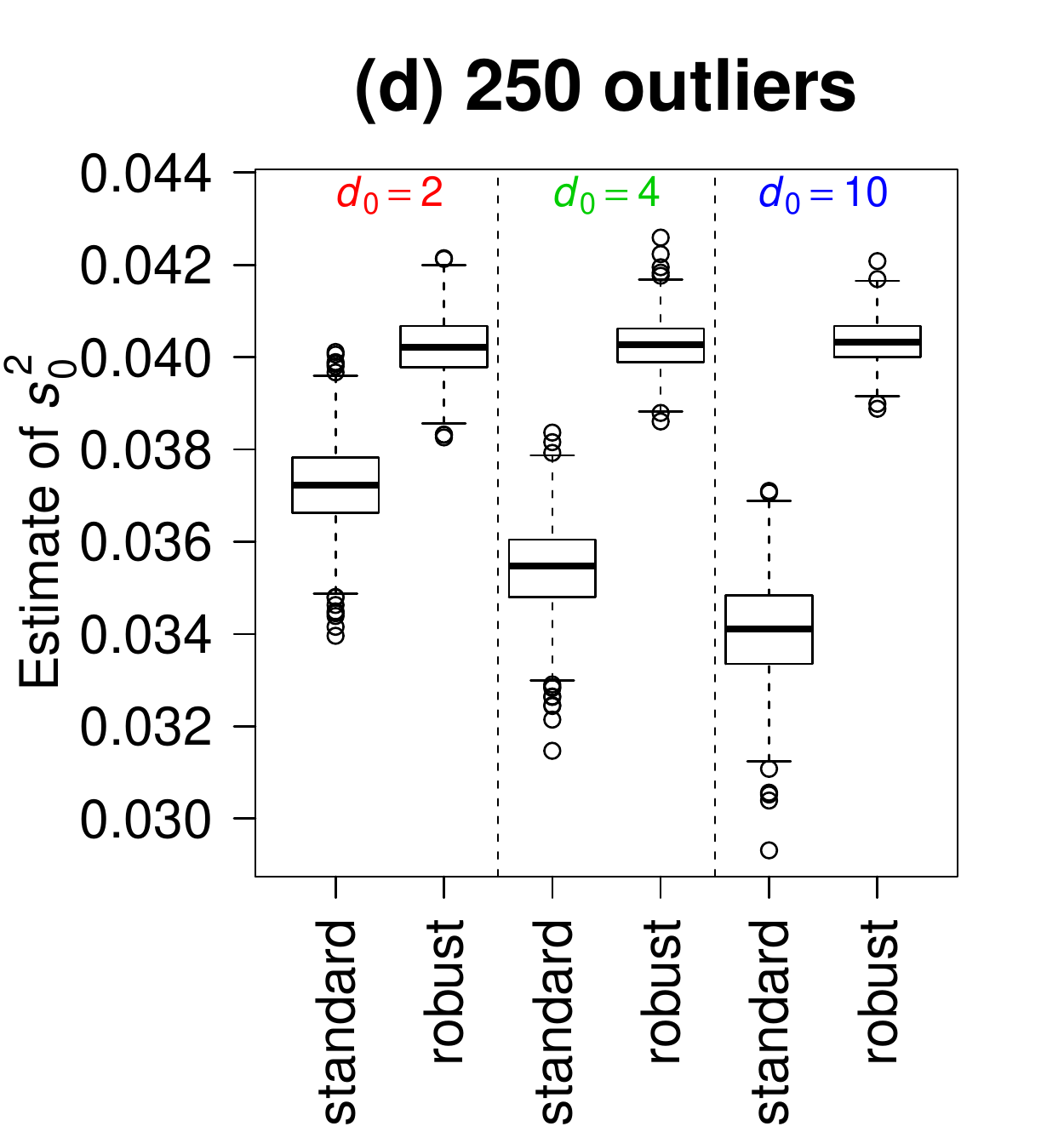}
\end{center}
\caption{Boxplots of standard and robust hyperparameter estimates from 1000 simulated datasets.
True values are 0.4 for $s_0^2$ and either 2, 4 or 10 for $d_0$.
Panels a, c show estimates when no outliers are present.
Panels b, d show estimates when the data includes 250 hypervariable genes.}
\label{hyperparamsim}
\end{figure}

\section{Evaluation using simulated data}
\label{section:sims}

Simulations were used to evaluate the performance of the robust hyperparameter estimators.
Expression values were generated for 10,000 genes and 6 RNA samples.
The RNA samples were assumed to belong to two groups, with three in each group, leading to a linear model with $d_g=4$ residual degrees of freedom.
Genewise variances and expression values were generated according to the hierarchical model of Section~2
with $s_0=0.2$ and with $d_0=2$, 4 or 10.
Both the standard and robust hyperparameter estimators were found to be accurate in the absence of outliers (Figure \ref{hyperparamsim}a, c).
When 250 hypervariable genes were included, however, the robust estimators were considerably more accurate than the standard (Figure \ref{hyperparamsim}b, d).
The hypervariable genes were simulated to have $d_{0g}=0.5$.
Next we checked type I error rates for the EB $t$-tests in the absence of outliers.
Both standard and robust tests were found to control the error rate correctly (Table \ref{typeIerror}).
In all simulations, the tail proportions $p_l$ and $p_u$ were at their default values.

\begin{table}
\caption{Type I error rates for standard and robust EB $t$-tests.
Datasets were simulated with different $d_0$ values but with no DE genes and or outliers.
The table gives the mean error rate over all genes in 1,000 simulated datasets for various $p$-value cutoffs.}
\label{typeIerror}
\begin{tabular}{rrllll}
      &        & \multicolumn{4}{c}{Nominal error rate} \\
\cline{3-6}
$d_0$ & Method & 0.001 & 0.01 & 0.05 & 0.1 \\
\hline
2  & Standard & 0.000996 & 0.00998 & 0.05001 & 0.09994 \rule{0pt}{12pt}\\
   & Robust & 0.000996 & 0.00998 & 0.04996 & 0.09983 \\
4  & Standard & 0.001008 & 0.01002 & 0.05008 & 0.10012 \rule{0pt}{12pt}\\
   & Robust & 0.001013 & 0.01004 & 0.05006 & 0.10005 \\
10 & Standard & 0.001017 & 0.01005 & 0.05018 & 0.10008 \rule{0pt}{12pt}\\
   & Robust & 0.001033 & 0.01010 & 0.05021 & 0.10005 \\
\end{tabular}
\end{table}

\begin{figure}
\begin{center}
\includegraphics[height=2.3in,width=2.2in]{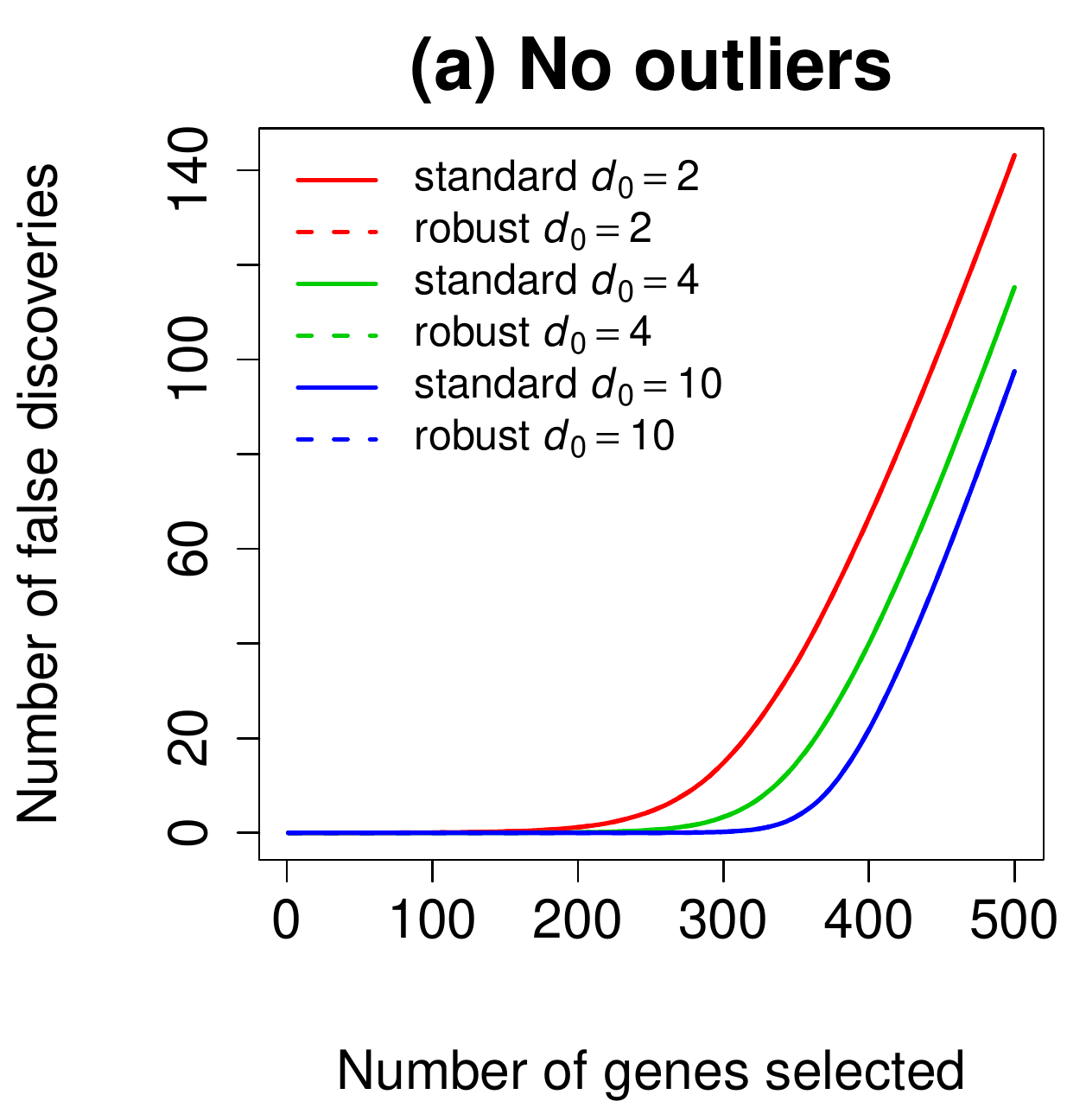}
\includegraphics[height=2.3in,width=2.2in]{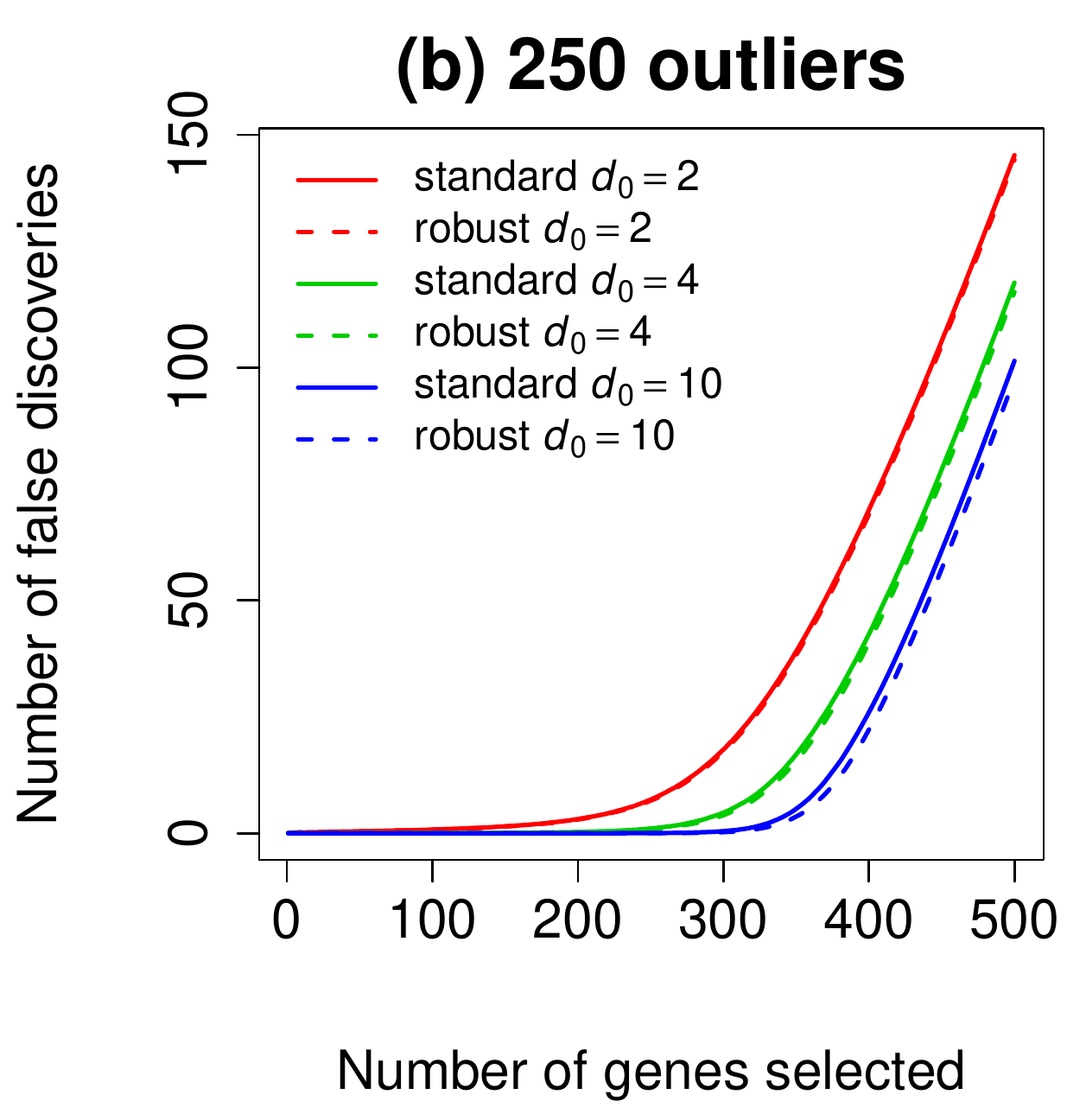}
\includegraphics[height=2.3in,width=2.2in]{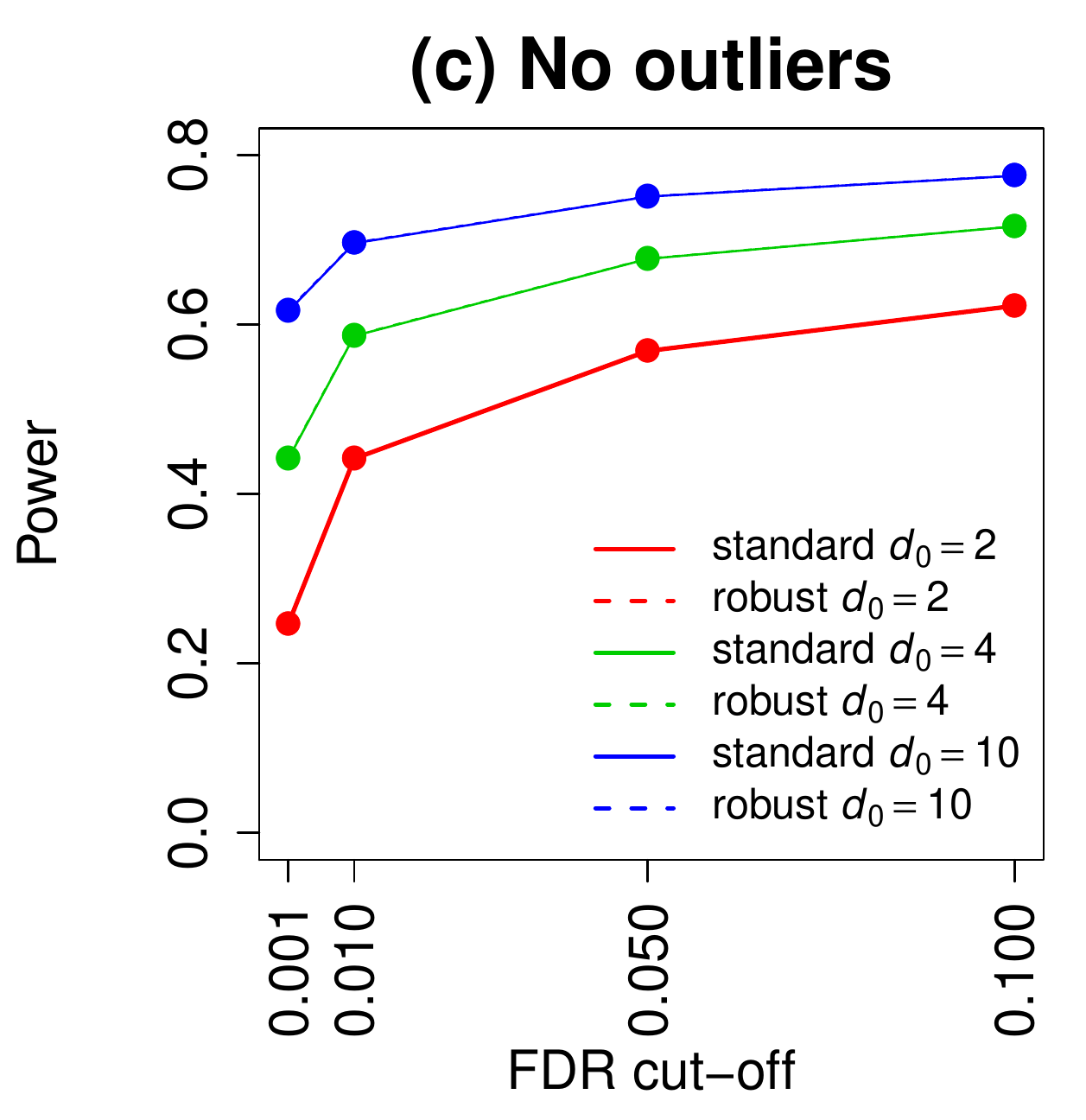}
\includegraphics[height=2.3in,width=2.2in]{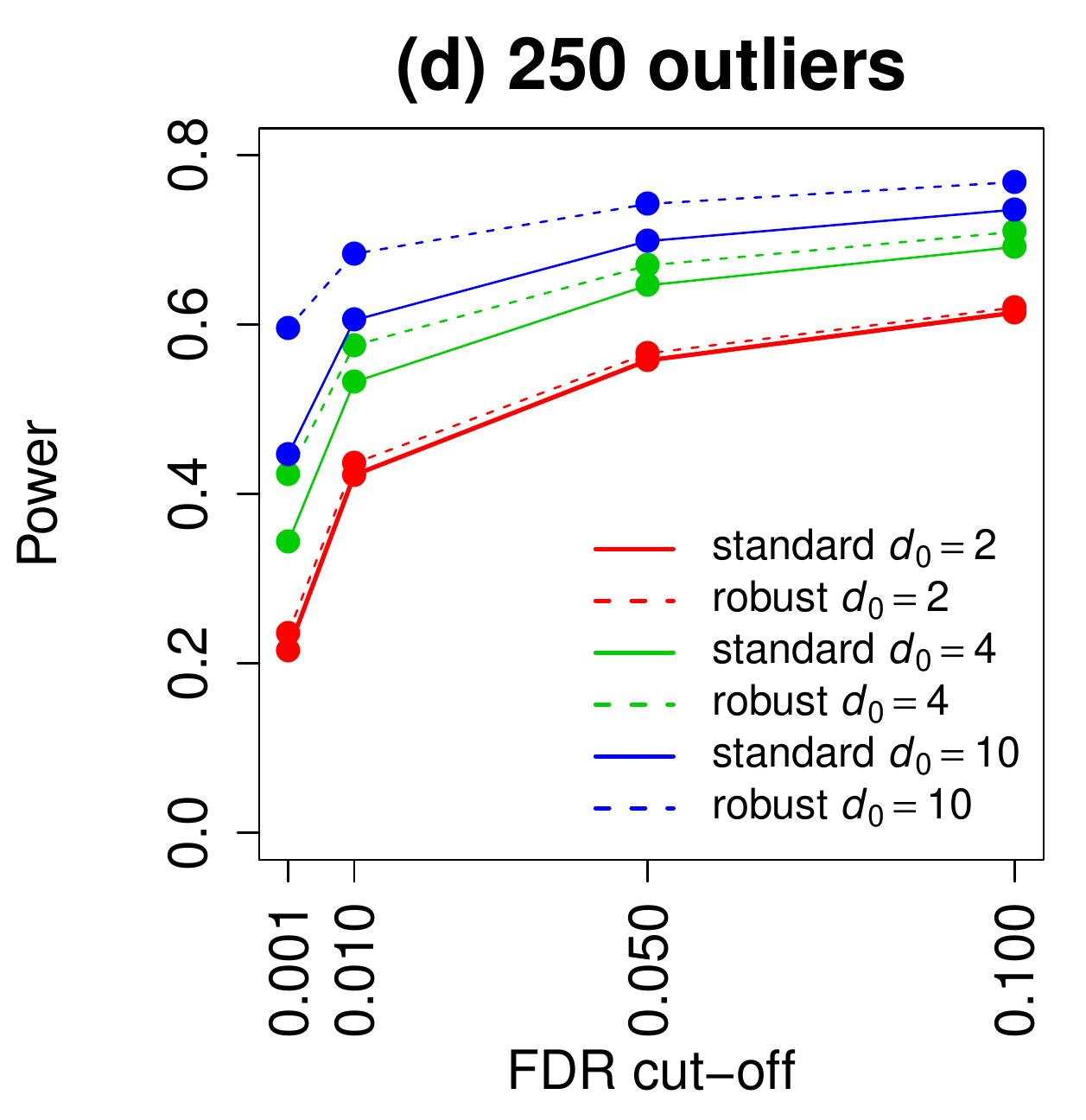}
\end{center}
\caption{Detection of differential expression.
Panels a--b show the number of false discoveries amongt the 500 top-ranked genes.
Panels c--d show power, the proportion of truly DE genes selected as significant at various FDR cutoffs.
Results are averaged over 1000 simulations.
Simulations b and d include 250 hypervariable genes.
Results are shown for both standard and robust EB tests and for three values of the prior degrees of freedom.}
\label{power}
\end{figure}

Finally we evaluated statistical power and FDR control.
Each simulated dataset now included 500 DE genes, with log fold changes generated from a $N(0,4)$ distribution.
In the absence of hypervariable genes, the standard and robust EB tests were indistinguishable in terms of false discoveries or power (Figure \ref{power}a,c).
In the presence of 250 outliers, the robust tests consistently gave fewer false discoveries (Figure \ref{power}b) and higher power (Figure \ref{power}d).
As expected, the improvement achieved by the robustified tests was greater for larger values of $d_0$.
For simplicity of interpretation, no genes were both DE and hypervariable in these simulations.

\section{Case studies}

\subsection{Loss of polycomb repressor complex 2 function in pro B cells}
\label{section:prc}

Polycomb group proteins are transcriptional repressors that play a central role in the establishment and maintenance of gene expression patterns during development.
Suz12 is a core component of Polycomb Repressive Complex 2 (PRC2).
\cite{majewski2008polycomb,majewski2010opposing} studied mice with a mutation in the \emph{Suz12} gene that results in loss of function of the Suz12 protein and hence PRC2.
They profiled gene expression in hematopoietic stem cells from these mice.
Here we describe a gene expression study of a different hematopoietic cell type from the same \emph{Suz12} mutant mice strain.
This study profiles gene expression in pro-B cells, an early progenitor immune cell intermediate in a series of development stages between hematopoietic stem cells and mature B-cells.

Our interest is to study development, so cells were isolated from 16-day embryonic mice.
For this study, RNA was extracted from foetal pro B cells that were isolated from the liver of four wild-type mice and four Suz12 mutant mice.
RNA was hybridized at the Australian Genome Research Facility to Illumina Mouse Whole-Genome-6 version 2 BeadChips, a microarray platform containing about 48,000 60-mer DNA sequences probing most genes in the genome.
Intensities were background corrected, quantile normalized and transformed to the $\log_2$-scale using the neqc function \citep{shi2010neqc}.
One of the \emph{Suz12} mutant samples was discarded because it clustered with the wildtype instead of the \emph{Suz12} samples, leaving four wildtype and three \emph{Suz12} mutant samples.
Probes were filtered from further analysis if they failed to achieve a detection p-value of less than 0.01 in at least two of the remaining samples, leaving 14,084 probes for analysis.

\begin{figure}[t]
\begin{center}
\includegraphics[height=2.5in,width=5.2in]{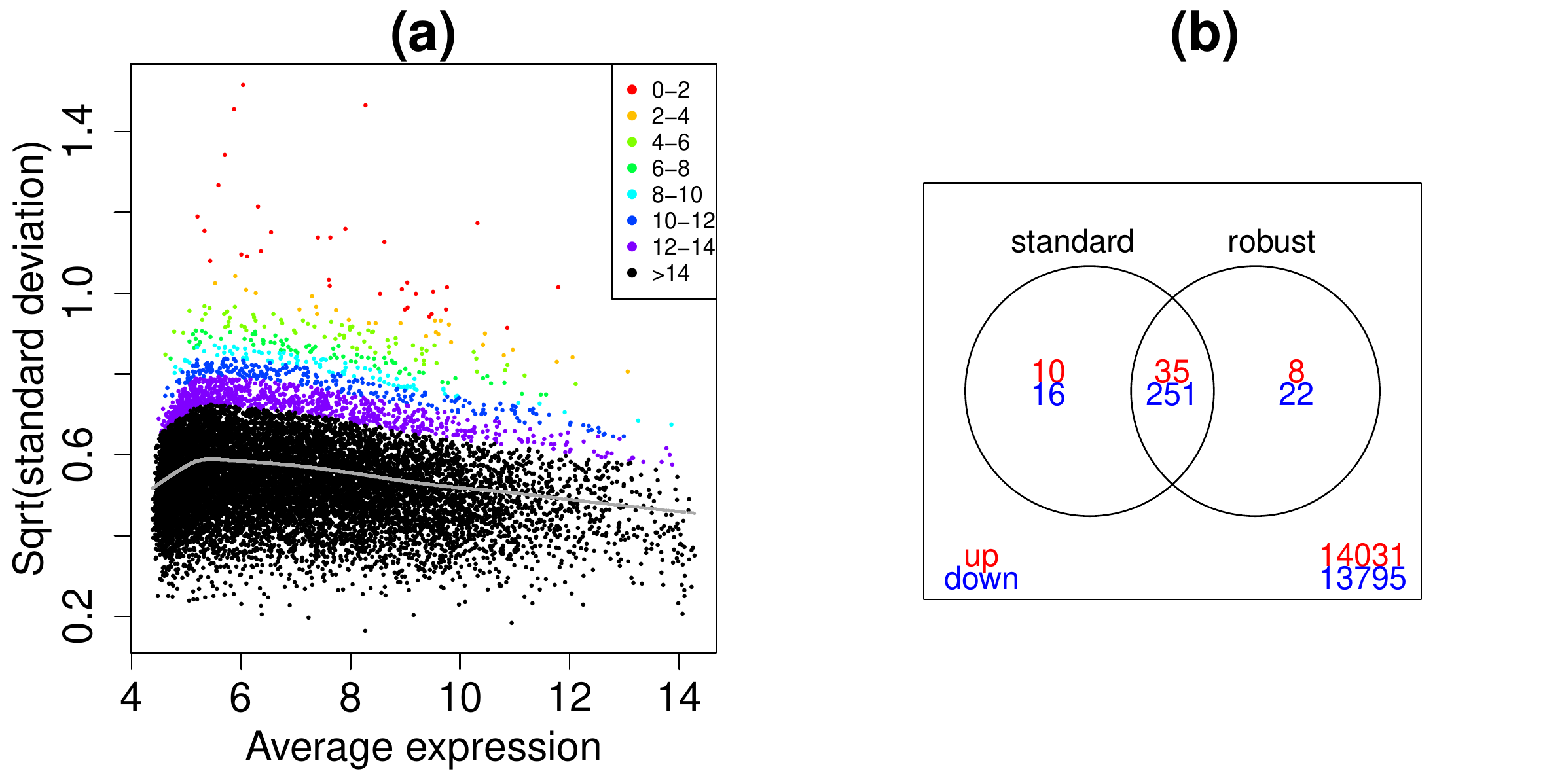}
\end{center}
\caption{
(a) Square-root standard deviation plotted against average log intensity for each probe.
The gray line shows the trended estimate of the prior variance.
Points are colored by the estimated prior degrees of freedom: probes with larger sample variances have smaller df.
(b) Venn diagram showing overlap the numbers of significant genes for Suz12 versus wildtype using the standard and robust methods.
}
\label{Fig4}
\end{figure}

Linear modeling was applied to the normalized log-expression values,
resulting in a residual sample variance on 5 degrees of freedom for each probe.
Figure \ref{PlotsShowingOutliers}a shows the residual standard deviation plotted against the average log intensity for each probe.
The gray curve shows the estimated trend for the prior variance.
The robust algorithm identified a number of outlier variances (Figure \ref{PlotsShowingOutliers}b).
The non-robust estimate of the prior degrees of freedom was 11.9.
The robust algorithm estimated prior degrees of freedom 14.1 for most genes, but with prior degrees of freedom as low as 0.5 for the outlier variances (Figure \ref{Fig4}a).

Further examination showed that many of the probes identified as outliers corresponded to genes known to have sex-linked expression, including many on the X or Y chromosomes (Figure \ref{PlotsShowingOutliers}ab).
The most outlying variances corresponded to Y chromosome genes \emph{Erdr1} and \emph{Eif2s3y} up-regulated in males, and X chromosome gene \emph{Xist}, known to be up-regulated in females.
Other outliers gene were ribosomal genes \emph{Rn18s} and \emph{Rpl7a}, suggesting ribosomal RNA retention in one or more samples, and hemoglobin genes \emph{Hbb-y} and \emph{Hbb-b1} suggesting red blood or bone marrow content in some tissue samples.
None of these genes should be related to the \emph{Suz12} mutation.

Differential expression between the \emph{Suz12} mutants and the wildtype mice was assessed using EB moderated $t$-statistics.
$P$-values were adjusted to control the false discovery rate at less than 5\% \citep{benjamini1995fdr}.
The standard and robust procedures found 251 down-regulated and 35 up-regulated probes in common (Figure \ref{Fig4}b).
However 22 and 16 down-regulated genes were found only by the robust or standard procedures respectively.
The non-robust unique genes tended to be sex-linked or hemoglobin related (\emph{Xist}, \emph{Apoa2}, \emph{Hbb-b1} etc) whereas the robust unique genes were related to programmed cell death (\emph{Bcl2l1}), cell cycle (\emph{Ccne2}) or chromatin remodeling (\emph{Myst2}).
For up-regulated genes, 8 and 10 unique probes were found by the robust and non-robust procedures respectively.
The non-robust unique genes tended to be Y chromosome sex-linked genes (\emph{Ddx2y}, \emph{Erdr1} etc) whereas the robust unique genes appeared related to the PRC2 process of interest.

Investigation after the analysis confirmed that two of the \emph{Suz12} mutant embryos were in fact female, whereas all the other mice were male.
This sex imbalance was an unwanted complication in the experiment, difficult to avoid without sex-typing of the embryo mice at the time of tissue collection.
The results show that the robust EB method was successful in identifying and downweighting genes that are associated with the hidden covariate.
The robust procedure results in more statistical power to detect other genes that are more likely to be of scientific significance.

\subsection{RNA-seq profiles of Yoruba HapMap individuals}

As part of the International HapMap project, RNA-Seq profiles were made of cell lines derived from B lymphocytes from 69 different Yoruba individuals from Ibadan, Nigeria \citep{pickrell2010understanding,pickrell2010noisy}.
Genewise read counts were obtained from the tweeDEseqCountData package version 1.8.0 from Bioconductor and were transformed to log2-counts per million with precision weights using voom \citep{law2014voom}.
The analysis compares males to females.
Figure~\ref{PlotsShowingOutliers}c shows the genewise standard deviations and Figure~\ref{PlotsShowingOutliers}d gives a probability plot of the standard deviations against the fitted $F_{d,d_0}$ distribution.
The hypervariable genes were identified as B cell receptor segments on chromosomes 2 and 22.
B cells contain a random selection of these segments in order to generate a repertoire of antigen binding sites.
The robust analysis reveals that the cell lines were clonal, each cell line apparently derived from a single B cell or from a very small number of original cells.
It would be appropriate to remove the receptor segments from the RNA-seq analysis, and the robust EB procedure effectively achieves that.

\subsection{Embryonic stem cells}

\cite{sheikh2015mozbmi1} used RNA-seq to profile embryonic stem cells.
Genewise read counts were transformed using voom.
Figure~\ref{PlotsShowingOutliers}e shows the genewise standard deviations and Figure~\ref{PlotsShowingOutliers}f gives a probability plot of the standard deviations against the fitted $F_{d,d_0}$ distribution.
The hyper-variable genes are predominately associated with the ribosome or with hemoglobin, suggesting inconsistenciees in cell purification and RNA processing.
Other hyper-variable genes are located in the major histocompatibility complex, known to be one of the most variable parts of the genome.

\section{Discussion}
\label{section:discussion}

In recent years we have routinely checked for hypervariable genes in gene expression studies.
We have found that many studies harbor a subset of outlier genes.
In many cases, the identities of the hypervariable genes suggest a mechanism for their variability.
We have analyzed studies, for example, where the hypervariable genes are enriched for sex-linked genes,
for ribosomal genes,
for mitochondrial genes,
or for B cell receptor segments.
In other cases, hypervariable genes may be associated with a particular cell type suggesting inconsistent cell population proportions in the different biological samples.
In some cases the reasons why some genes are highly variable between individuals are unknown.
The phenomenon is common enough to be viewed as an unavoidable part of cutting edge genomic research rather than a result of flawed experimental procedures.
In most cases the studies are overall high quality.

Hypovariable genes can also arise, although usually for technical rather than biological reasons.
Quantile normalization \citep{bolstad2003comparison} of expression data can occasionally produce expression values that are numerically identical for all samples for a given gene when the number of samples is small.
Sequencing data can also give rise to variances that are zero or very small because of the discreteness of read counts.

This article describes a robustified version of EB differential expression analysis.
This procedure protects against hyper and hypovariable genes in the sense that it allows non-outlier genes to share information amongst themselves as if the outlier genes were not present.
In many cases, this results in a gain in statistical power for the non-outlier genes.
Hypervariable genes are not removed from the analysis but instead borrow less information from the ensemble and are assigned test statistics closer to ordinary $t$-statistics.
Hypovariable genes on the other hand are moderated as for non-outlier genes---this increases their posterior variances closer to typical values.

A key feature of our procedure is that a conjugate Bayesian model is used for each gene,
enabling closed-form posterior estimators and exact small-sample $p$-values.
Robustness is achieved by fitting the prior distribution to non-outlier genes and by assigning lower degrees of freedom to the hypervariable outliers.
We have proposed a practical algorithm for assigning prior degrees of freedom to outlier genes.
In practice, the list of DE genes is not sensitive to the exact values of the prior degrees of freedom, provided that $d_{0g}=d_0$ for non-outliers and $d_{0g}$ is substantially smaller for clear outliers.
For many datasets the list of DE genes is unchanged for a range of reasonable values for $d_{\rm outlier}$.

The default values for the Winsorizing tail proportions $p_l$ and $p_u$ work well in our practical experience.
Users however may choose to increase the default values for datasets where high proportions of outlier genes are expected.

Simulations show that the robustified EB procedure estimates the hyperparameters equally as accurately as the original method in the absence of outliers.
When outliers are present, however, the robustified EB procedure was able to simultaneously increase power and decrease the false discovery rate when assessing differential expression.

The robust EB method developed here has been applied not only to microarray data, but also to data from RNA-seq \citep{goodjacobson2014moz}, ChIP-seq \citep{lun2015csaw,lun2015csawworkflow} and Hi-C \citep{lun2015diffhic} technologies.
With microarrays, the linear models are fitted to normalized log-intensities.
With RNA-seq, the number of sequence reads overlapping each gene can be counted and the EB models can be applied to the log-counts-per million \citep{law2014voom}.
Alternatively, the robust EB method can be applied to count data by way of quasi-generalized linear models.
\cite{lun2016delicious} analyzed RNA-seq read counts used quasi-negative-binomial generalized linear models.
The limma robust EB variance estimation method was applied to the genewise residual deviances leading to the construction of EB quasi-F-tests.
The robust EB method has also been used to estimate the prior degrees of freedom for the weighted likelihood approach used in the edgeR package, again using residual deviances \citep{chen2014edgerchapter}.

In this article we view genes as outliers rather than individual expression values as outliers.
An alternative robustifying approach would be to replace least squares estimation of the genewise linear models with robust regression \citep{gottardo2006bayesian,zhou2014robustly}, and the limma package has included an M-estimation option for this purpose for over a decade.
In principle, the two approaches are complementary and both can be used simultaneously, i.e., one could apply the robust EB procedure of this article to variances estimated by robust regression.
The robust regression approach assumes that the expression values for a gene contain one or two outliers that are ``errors'' whereas the other values are ``correct''.
Our experience suggests that this scenario is relatively rare for gene expression data.
The outlier-gene approach of this article allows a more general context in which a hypervariable gene may produce an arbitrary number of inconsistent expression values that cannot be meaningfully categorized into correct and incorrect.
The robust regression is only applicable to experimental designs with at least three expression values per experimental group, whereas the outlier-gene approach can be usefully applied to any experimental design, even down to studies with a single residual degree of freedom.

The robust EB strategy of this article could in principle be applied in other EB contexts.
The basic idea is to estimate hyperparameters robustly, then to test for outlier cases, and finally to assign a more diffuse prior to outlier cases.
This approach may be particularly attractive for use with conjugate Bayesian models and seems different from previous robust EB strategies.

It is interesting to contrast our approach with the large literature on robust Bayesian analysis \citep{berger1984robustbayesianviewpoint,berger1990robustbayesiananalysis,insua2000robustbayesiananalysis}.
Robust Bayesian analysis considers a class of possible prior distributions and tries to limit or at least quantify the range of posterior conclusions as the prior ranges over the class.
The issues that concern us in this article are different in a number of important ways.
The issues that we address are specific to empirical Bayes and do not arise in true Bayesian frameworks for which hyperparameters do not need to be estimated.
Our aim is not to limit the influence of the prior but to increase it for the majority of genes.
Our method applies only to large scale data with many cases (probes, genes or genomic regions) whereas robust bayesian analysis is typically concerned with individual cases.

\section{Appendix: Details of computational algorithms}
\label{supp}

\subsection{Standard (non-robust) moment estimation of the hyperparameters}

In order to calculate the EB moderated $t$-statistics, the hyperparameters $d_0$ and $s_0^2$ need to be estimated.
Here we summarize the original (non-robust) methods-of-moments strategy for estimating the hyperparameters proposed by \cite{smyth2004ebayes}.
The marginal distribution of the gene-wise residual variances $s^2_g$ is $s_0^2 F_{d_g,d_0}$.
It is convenient to work with the logged variances, defined by $z_g=\log s_g^2$, as these have finite moments and a roughly normal distribution for which the methods of moments should be efficient.
The theoretical mean and variance of $z_g$ are available as
\[
E(z_g) = \log s_0^2 + \psi\left(\frac{d_g}{2}\right) -\psi\left(\frac{d_0}{2}\right)+\log\left(\frac{d_0}{d_g}\right)
\]
and
\[
\textrm{var}(z_g) = \psi'\left(\frac{d_g}{2}\right)+\psi'\left(\frac{d_0}{2}\right),
\]
where $\psi$ and $\psi'$ represent the digamma and trigamma functions respectively.
Write $\bar z_g$ and $s^2_z$ for the observed mean and variance of the $z_g$.
These have theoretical expections
\[
E(\bar z_g) = \log s_0^2 -\psi\left(\frac{d_0}{2}\right) +\log\left(\frac{d_0}{2}\right) + \frac1G\sum_{g=1}^G \left\{\psi\left(\frac{d_g}{2}\right) - \log\left(\frac{d_g}{2}\right) \right\}
\]
and
\[
E(s^2_z) = \psi'\left(\frac{d_0}{2}\right) + \frac1G\sum_{g=1}^G \psi'\left(\frac{d_g}{2}\right)
\]
The equation for $E(s^2_z)$ can be solved for $d_0$, then that for $E(\bar z_g)$ can be solved for $s^2_0$.

Alternatively a abundance-dependent trend can be put on $s_0$.
In that case, a regression spline with 4 df is used to model $z_g-\psi(d_g/2)+\log(d_g/2)$ a function of average $\log_2$-expression, $\bar y_g$, and the fitted values from the regression are used in place of $\bar z_g-\psi(d_g/2)+\log(d_g/2)$.

\subsection{Tranforming samples variances to equal residual degrees of freedom}

The ordinary non-robust moment estimators for the hyperparemeters do not require the residual degrees of freedom to be equal for all genes, but the robust estimation scheme does require this.
In practice, the $d_g$ are usually equal, but occasionally some genes may have reduced $d_g$ because of missing expression values for some samples.
If this is the case, we transform the sample variances $s^2_g$ so they can be treated as being on the same degrees of freedom.
Let $d$ be the maximum $d_g$.
First the hyperparameters $d_0$ and $s^2_0$ are estimated by the non-robust algorithm.
Then the $s^2_g$ are transformed to $s^2_0 F^{-1}_{d,d_0} F_{d_g,d_0}(s^2_g/s^2_0)$ where $F_{k_1,k_2}$ denotes here the cumulative distribution function of the $F$-distribution on $k_1$ and $k_2$ degrees of freedom.

\subsection{Computing Winsorized moments}

Here we describe the computation of the mean $\nu(d_g,d_0)$ and variance $\phi(d_g,d_0)$ of the log Winsorized $F$-distribution.
Let $Z=\log {\rm win}(f)$ denote the log-Winsorized $F$ random variable defined in Section 3 of the main paper.
The expected value is
$$\nu(d_g,d_0)=E(Z)=p_l\,\log q_l+p_{lu}\, E\left(Z\,|\,q_l<Z<q_u\right)+p_u\,\log q_u$$
with $p_{lu}=(1-p_l-p_u)$.
After transforming $Z$ to the unit interval,
the conditional expectation can be re-interpretted as
\[
p_{lu}\,E(Z|q_l<Z<q_u)=(b-a)\,E\{h(U)\}
\]
where $U$ is uniformly distributed on the interval from $a=q_l/(1+q_l)$ to  $b=q_u/(1+q_u)$ and
\[
h(u) = \log \left(\frac{u}{1-u} \right) \frac{1}{(1-u)^2}\,{\rm pdf}\left(\frac{u}{1-u}\right)
\]
where pdf is the probability density function of the $F$-distribution on $d_g$ and $d_0$ degrees of freedom.
The expectation in terms of the uniform random variable can be evaluated efficiently by Gauss-Legendre quadrature, as described in the next section.

Having computed the mean, the variance of $Z$ can be obtained as
\begin{eqnarray*}
\phi(d_g,d_0)&=&E\left\{(Z-\nu)^2\right\}\\
&=&p_l\,(\log q_l-\nu)^2+p_{lu}\, E\left\{(Z-\nu)^2\,|\,q_l<Z<q_u\right\}\\
&&+p_u\,(\log q_u-\nu)^2
\end{eqnarray*}
where $\nu=\nu(d_g,d_0)$.
The second term containing the conditional expectation can be re-interpretted as $(b-a)\,E\{h(U)\}$ with
\[
h(u) = \left\{\log \left(\frac{u}{1-u} \right)-\nu\right\}^2 \frac{1}{(1-u)^2}\,{\rm pdf}\left(\frac{u}{1-u}\right)
\]
to permit evaluation by Gauss-Legendre quadrature.

\subsection{Evaluating an integral using Gaussian quadrature}

To compute the Winzorized moments, we need $E\{h(U)\}$ where $U$ is uniformly distributed on the interval $[a,b]$.
We compute this expectation using Gauss-Legendre quadrature.
Specifically, we use the function \texttt{gauss.quad.prob} from the statmod package \citep{smyth2015statmod}.
This function computes Gauss quadrature weights and nodes using an adaption of the algorithm and Fortran code published by \cite{golub1969calculation}.
For any desired order $k$, nodes $u_i$ and weights $w_i$ can be computed such that
\[
E \{ h(U)\} \approx \sum_{i=1}^k w_i h(u_i).
\]
The approximation is exact if $h(u)$ can be expressed as a polynomial of order $2k-1$ or less on $[a,b]$.
The accuracy of Gauss-Legendre quadration is excellent if $h(u)$ is a reasonably smooth function taking finite values on the interval.

All results reported in the article use $k=128$ nodes.
This is sufficient for close to double-precision accuracy provided $a$ is bounded above zero and $b$ is bounded below one.

\subsection{Solving for $d_{\rm outlier}$}

Let $s^2_{\rm max}$ be the maximum of $s^2_g/s^2_{0g}$ over all genes $g$.
We wish to find $d_{\rm outlier}$ such that $\bar F(s^2_{\rm max};d_g,d_{\rm outlier})=0.5$, where $\bar F$ is the right tail probability of the F distribution with $d_g$ and $d_0$ degrees of freedom.
Initializing $d_{\rm outlier}=d_0$ and then repeating
\begin{eqnarray*}
p&=&\bar F(s^2_{\rm max};d_g,d_{\rm outlier})\\
d_{\rm outlier}&=&d_{\rm outlier} \times \log(0.5)/\log p 
\end{eqnarray*}
converges monotonically to the required value.
Sufficient accuracy is achieved after two or three iterations.

\section*{Acknowledgments}

This work was supported by The University of Melbourne (PhD scholarship to BP),
by the National Health and Medical Research Council (Fellowship 1058892, Program Grant 1054618, Independent Research Institutes Infrastructure Support Scheme),
and by a Victorian State Government OIS grant.

\end{document}